 \documentclass[]{jfm}

\usepackage{graphicx}
\usepackage{amsmath}%
\usepackage{subfig}
\usepackage{psfrag}
\usepackage[usenames, dvipsnames]{color}
\usepackage{float}
\usepackage{natbib}
\usepackage{pifont}
\usepackage{ar}

\ifCUPmtlplainloaded \else
  \checkfont{eurm10}
  \iffontfound
    \IfFileExists{upmath.sty}
      {\typeout{^^JFound AMS Euler Roman fonts on the system,
                   using the 'upmath' package.^^J}%
       \usepackage{upmath}}
      {\typeout{^^JFound AMS Euler Roman fonts on the system, but you
                   dont seem to have the}%
       \typeout{'upmath' package installed. JFM.cls can take advantage
                 of these fonts,^^Jif you use 'upmath' package.^^J}%
      }
  \else
  \fi
\fi

\ifCUPmtlplainloaded \else
  \checkfont{msam10}
  \iffontfound
    \IfFileExists{amssymb.sty}
      {\typeout{^^JFound AMS Symbol fonts on the system, using the
                'amssymb' package.^^J}%
       \usepackage{amssymb}%
       \let\le=\leqslant  \let\leq=\leqslant
       \let\ge=\geqslant  
      }{}
  \fi
\fi

\ifCUPmtlplainloaded \else
  \IfFileExists{amsbsy.sty}
    {\typeout{^^JFound the 'amsbsy' package on the system, using it.^^J}%
     \usepackage{amsbsy}}
    {}
\fi

\newcommand\Rey{\mbox{\textit{Re}}}  

\newsavebox{\astrutbox}
\sbox{\astrutbox}{\rule[-5pt]{0pt}{20pt}}

\newcommand{\diff}{\mathrm{d}}

 \title[]{On turbulent friction in straight ducts with complex cross section: the wall law and the hydraulic diameter}

\author[S. Pirozzoli]{SERGIO PIROZZOLI}

\affiliation{Dipartimento di Ingegneria Meccanica e Aerospaziale,
Sapienza Universit\`a di Roma \\ Via Eudossiana 18, 00184 Roma,
Italy \\[\affilskip]}

\pubyear{1996}
\volume{538}
\pagerange{119--126}

\begin{document}

\maketitle

\begin{abstract}
We develop predictive formulas for friction resistance in ducts with complex cross-sectional shape based on the use of the log law and neglect of wall shear stress nonuniformities. The traditional hydraulic diameter naturally emerges from the analysis as the controlling length scale for common duct shapes as triangles and regular polygons. The analysis also suggests that a new
effective diameter should be used in more general cases, yielding corrections 
of a few percent to friction estimates based on the traditional hydraulic diameter.
Fair but consistent predictive improvement is shown for duct geometries of practical relevance, including rectangular and annular ducts, and circular rod bundles.
\end{abstract}

\section{Introduction}\label{sec:intro}

Internal flows within straight ducts having non-circular cross section 
are common in many applications of mechanical and hydraulic engineering, 
including water draining and ventilation systems, nuclear reactors,
heat exchangers and turbomachinery.
Despite their practical importance, flows in ``complex'' ducts are understood to a much lesser extent compared to canonical flows in plane channels and circular pipes, one
of the main qualitative differences being the occurrence of 
secondary motions~\citep{prandtl_27,nikuradse_30}.
Of primary engineering importance is the prediction of the duct friction factor, namely
\begin{equation}
\lambda = \frac {8 \overline{\tau}_w}{\rho u_b^2}, \label{eq:Cf}
\end{equation}
where $\overline{\tau}_w$ is the average wall shear stress along the duct perimeter,
$\rho$ is the fluid density, and $u_b$ is the duct bulk velocity.
For that purpose, a widely used pragmatic approach is assuming that the same friction relationship
can be used as for flow in circular pipes, provided a suitable Reynolds number is defined, based 
on a convenient duct length scale. 
The most traditional choice, apparently initiated by \citet{schiller_23}, is the 
hydraulic diameter, defined as
\begin{equation}
D_h = \frac{4 A}{P_0}, \label{eq:DH}
\end{equation}
where $A$ is the duct cross-sectional area, and $P_0$ is the duct perimeter.
This choice is frequently justified by the appearance of the area/perimeter ratio in the mean momentum
balance equations, which connects the mean wall friction to the duct pressure drop, namely
\begin{equation}
\overline{\tau}_w = - \frac AP_0 \frac{\diff p}{\diff x}, \label{eq:mmb}
\end{equation}
but of course this is a quite tenuous argument. Indeed, limitations of the 
hydraulic diameter have emerged, especially in the study of ducts with aspect ratio
(here broadly defined as the ratio of the largest to the smallest dimensions of the duct)
significantly different than unity. Hence, semi-empirical corrections to the basic hydraulic diameter 
concept have been proposed over the years, tailored to specific duct geometries, 
as for instance rectangular ducts~\citep{jones_76}, 
and general geometries have been frequently handled 
through ad-hoc extrapolation of laminar results~\citep[e.g.][]{maubach_70, rehme_72}.
More elaborate techniques have sometimes been used relying on application of the log law
in direction normal to the velocity iso-lines, which however require a cumbersome
iterative procedure~\citep{deissler_58}.

One of the most robust findings in the (few) detailed quantitative studies of non-canonical
duct flow is the presence of logarithmic layers in the wall-normal direction~\citep[e.g.][]{cain_71,nouri_93,jonsson_66}.
Even more convincing support for the validity of the law-of-the-wall has come from 
recent numerical studies dealing with rectangular ducts~\citep{vinuesa_14}, square ducts~\citep{pirozzoli_18},
and hexagonal ducts~\citep{marin_16}, showing that in its inner form it applies with excellent
approximation in the wall-normal direction up to the nearest corner bisector, even near the
duct corners, provided the local wall friction is used. 
A formal theoretical analysis of turbulent flow in ducts with complex shape has been recently developed by \citet{spalart_18}. 
Extending classical inner-outer layer matching arguments, those authors 
deduced the validity of the logarithmic velocity profile, 
but also arrived at the interesting prediction that the wall friction
should tend towards being uniform all around the duct except near possible corners, 
asymptotically as $\Rey \to \infty$. This was confirmed by numerical solutions obtained
with turbulence models, and it is consistent 
with recent DNS at moderately high Reynolds number~\citep{pirozzoli_18}.
Hence, a reasonably simple structure of turbulent flow in ducts emerges, which lends itself
to analytical treatment. In the forthcoming \S\ref{sec:theory} we 
develop a simple predictive formula for the friction coefficient for ducts with arbitrary shape,
and in \S\ref{sec:applications} we test its predictive capability.
Last, in \S\ref{sec:concl} we outline implications of the present findings.

\section{Friction estimates}\label{sec:theory}

\begin{figure}
 \begin{center}
  \psfrag{y}[][][1.4]{$y$}
  \psfrag{p}[][][1.4]{$X$}
  \psfrag{pw}[][][1.4]{$X_w$}
  \psfrag{D}[][][1.4]{$\mathcal{D}$}
  \psfrag{Py}[][][1.4]{$P(y)$}
  \includegraphics[width=5.0cm,clip]{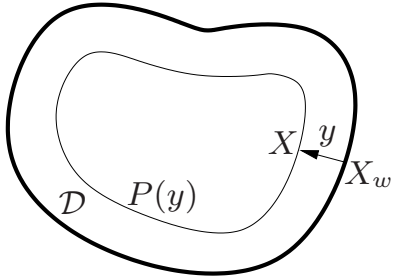}
\caption{Typical duct cross section ($\mathcal{D}$), with indication of wall distance $y$, and associated area density, $P(y)$. $X$ is a generic point and $X_w$ is its wall foot.}
  \label{fig:duct}
 \end{center}
\end{figure}

The forthcoming discussion pertains to straight ducts with cross-sectional shape $\mathcal{D}$ 
such that a wall distance $y$ (defined at the normal distance to the closest wall) can be defined for any point $X$, as depicted in figure~\ref{fig:duct}. This is a generally acceptable assumption, and counterexamples are difficult to conceive. We hereafter further assume that:
\begin{enumerate}
\item \label{it:1} the inner-layer law-of-the wall applies with respect to the closest wall,
hence 
\[ \overline{u}(X)/u_{\tau} = f(y u_{\tau} / \nu) , \] 
where $u_{\tau} = (\tau_w/\rho)^{1/2}$ is the local friction velocity at the wall point $X_w$,
and $f$ is a suitable functional inner form for the wall law;
\item \label{it:2} the wall shear stress is constant along the duct perimeter, hence $\tau_w = \overline{\tau}_w$.
\end{enumerate}
Assumption~(\ref{it:1}) is supported by all available experimental and numerical data~\citep{cain_71,jonsson_66,nouri_93,marin_16,nikitin_06}. 
For instance, in square ducts~\citep{pirozzoli_18}
clear logarithmic layers are found to form at bulk Reynolds number barely exceeding $10^4$.
Small flow-dependent deviations from the strict inner law are present in the core part of the
flow, as for instance in circular pipes and plane channels, which will be neglected here for convenience.
Assumption~(\ref{it:2}) is probably weaker, as direct detailed measurements of the local wall friction 
distribution are scarce, and numerical simulations are limited to relatively low Reynolds numbers.
The available data show that, with the obvious exception of sharp corners where the friction is zero, 
variations of the wall shear stress in simply-connected ducts are no more that $10\%$~\citep{leutheusser_63,cain_71,marin_16}.
In the best documented case of square ducts, variations of local wall friction are further 
found to very nearly cancel out as far as their contribution to the mean 
friction coefficient is concerned~\citep{pirozzoli_18}.
Compelling theoretical arguments~\citep{spalart_18} do in fact suggest that flat distributions of the wall
shear stress should result in the asymptotic high-Reynolds-number limit.
Larger variations of wall friction  have been reported in multiply-connected ducts, as discussed in \S\ref{sec:annular}.
Although certainly criticizable, the assumptions~(\ref{it:1}),(\ref{it:2}) are the only viable pathway
to arrive at closed form predictions, hence their validity may be judged from the outcome.
Similar assumptions were also used by \citet{keulegan_38} to estimate friction in open channels
with trapezoidal cross section, and are frequently used in the hydraulics community.
However, it appears that implications have not been fully pursued in the past,
especially within the context of internal flows. 

A direct consequence of the above assumptions is that the mean streamwise velocity only depends on
the inner-scaled wall distance according to
\begin{equation}
u^* = f ( {y^*} ), \label{eq:lotw}
\end{equation}
where the star denotes normalization with global wall units, namely $u^* = \overline{u}/u_{\tau}^*$,
$y^* = y u_{\tau}^* / \nu$, with $u_{\tau}^* = (\overline{\tau}_w/\rho)^{1/2}$.
The bulk velocity in the duct is then easily deduced by integration, yielding
\begin{equation}
u_b^* = \frac 1A \int_{\mathcal{D}} f(y^*) \diff A. \label{eq:ub1}
\end{equation}
The integration is conveniently carried out by introducing the area density with respect to the 
wall distance, $P(y) = \diff A / \diff y$ (see figure~\ref{fig:duct}), which may be interpreted as the perimeter associated 
with the iso-$y$ lines, with $P(0)=P_0$, thus obtaining
\begin{equation}
u_b^* = \frac 1A \int_0^{y_m} P(y) f(y^*) \diff y = \int_0^1 \tilde{P} (\eta) f(\eta \cdot y_m^*) \diff \eta, \label{eq:ub2}
\end{equation}
where $y_m$ is the maximum wall distance over the duct cross section, $\eta = y/y_m$ is the 
normalized wall distance, $y_m^* = y_m u_{\tau}^* / \nu$ is the counterpart of the friction Reynolds number in canonical flows,
and $\tilde{P} (\eta) = P(\eta y_m)/\overline{P}$ is the normalized perimeter function, with $\overline{P} = A/y_m$ the mean duct perimeter. Equation~\eqref{eq:ub2} makes it clear that the wall-scaled bulk velocity, and hence the friction factor (from equation~\eqref{eq:Cf}, $\lambda = 8/{u_b^*}^2$), only depends
upon the wall function, which is here assumed to be the same for all ducts, and on the duct geometry through 
the inner-scaled maximum distance $y_m^*$, and the normalized perimeter function.
For instance, in the case of a circular pipe with diameter $D$ it is straightforward to verify that $y_m=D/2$, $\overline{P} = \pi D/2$, and $\tilde{P} = 2 (1-\eta)$. Notably, very similar conclusions are 
obtained for any triangle and for regular polygons, in which case $y_m$ is the apothem of the cross section,
$\overline{P}$ is half of the outer perimeter, and again $\tilde{P} = 2 (1-\eta)$. This observation leads to the 
first important conclusion that, based on the previous assumptions, 
triangles and regular polygons should share the same friction relation, provided
the apothem is used as length scale in the definition of the Reynolds number. 
However, it is easy to show that for all these geometries the hydraulic diameter defined in equation~\eqref{eq:DH} 
is two times the duct apothem,  
hence we find that the assumed strict validity of the wall law yields as a direct, exact consequence, 
that the hydraulic diameter is the proper length scale to achieve universality of the
friction coefficient distribution.

Additional elaborations can be made by assuming a logarithmic form 
for the law-of-the-wall which formally encompasses both the case of smooth and rough walls, namely
\begin{equation}
u^* = \frac 1k \log (y^* / y_0^*) = \frac 1k \log (y/y_0), \label{eq:lotwg}
\end{equation}
where $k$ is the von Karman constant, and $y_0$ is the virtual origin for the wall law, defined as~\citep{colebrook_39}
\begin{equation}
y_0 = 
\frac {\alpha \nu}{u_{\tau}^*} + 
\beta \varepsilon,
\end{equation}
where $\varepsilon$ is the equivalent sand-grain roughness height, and $\alpha \approx 1/10$, $\beta \approx 1/33$.
Partial integration of~\eqref{eq:ub2} yields
\begin{equation}
u_b^* = \frac 1k \log \left( y_m / y_0 \right) + \frac 1k \underbrace{\int_0^{1} \tilde{P}(\eta) \log \eta \, \diff \eta}_{\mathcal{C}} , \label{eq:ub3}
\end{equation}
where $\mathcal{C}$ is solely a function of the duct geometry.
The same asymptotic Reynolds number trend also directly stems from 
matching arguments~\citep{spalart_18}, however the analysis is here completed by
integration over the cross section, thus yielding full friction predictions.

In the case of flow in a circular pipe, Eqn.~\eqref{eq:ub3} becomes
\begin{equation}
u_b^* = \frac 1k \log (D/2/y_0)- \frac 3{2k} . \label{eq:ub3_pipe}
\end{equation}
Comparing equations~\eqref{eq:ub3} and \eqref{eq:ub3_pipe} shows that the two 
formulas are identical provided $D$ in equation~\eqref{eq:ub3_pipe} is replaced with
\begin{equation}
D_e = 2 y_m e^{3/2 + \mathcal{C}} , \label{eq:effdiam}
\end{equation}
which may then be defined as an effective diameter for the duct, namely 
the diameter of a circular pipe yielding the same friction coefficient.
As previously pointed out, the effective diameter herein predicted coincides
with the traditional hydraulic diameter in ducts with triangular and regular polygonal shape, for which $\mathcal{C}=-3/2$.
Equation~\eqref{eq:effdiam} however highlights that the correct length scale to achieve 
universality of the friction law is in general different than the hydraulic diameter. 
Differences in the prediction of $C_f$ may be estimated considering the extreme case
of an infinitely wide channel with height $H$, whose hydraulic diameter is $D_h=2H$,
and whose normalized perimeter function is $\tilde{P}\equiv 1$. 
From equation~\eqref{eq:effdiam} it follows 
that $D_e = \sqrt{e} H \approx 1.65 H$,
which by construction returns the log-law based friction law for plane channel flow.
Assuming for simplicity $\lambda \sim \Rey_D^{-1/4}$~\citep{blasius_13}, we estimate that
use of the hydraulic diameter in this case yields underestimation of the
friction coefficient of about $5\%$, as indeed confirmed by the later results.

In the following we verify the predictive power of 
the traditional hydraulic diameter and of the effective diameter~\eqref{eq:effdiam}
for smooth ducts with relatively complex geometry, and for which a sufficient number of experimental 
measurements are available.

\section{Applications} \label{sec:applications}

\subsection{Rectangular ducts} \label{sec:rectangle}

\begin{figure}
 \begin{center}
  \psfrag{y}[t][][1.2]{$y$}
  \psfrag{p}[b][][1.2]{$X$}
  \psfrag{pw}[b][][1.2]{$X_w$}
  \psfrag{D}[t][][1.2]{$D$}
  \psfrag{Py}[][][1.2]{$P(y)$}
  \psfrag{2a}[][][1.2]{$2a$}
  \psfrag{2b}[][][1.2]{$2b$}
  (a)
  \includegraphics[width=6.5cm,clip]{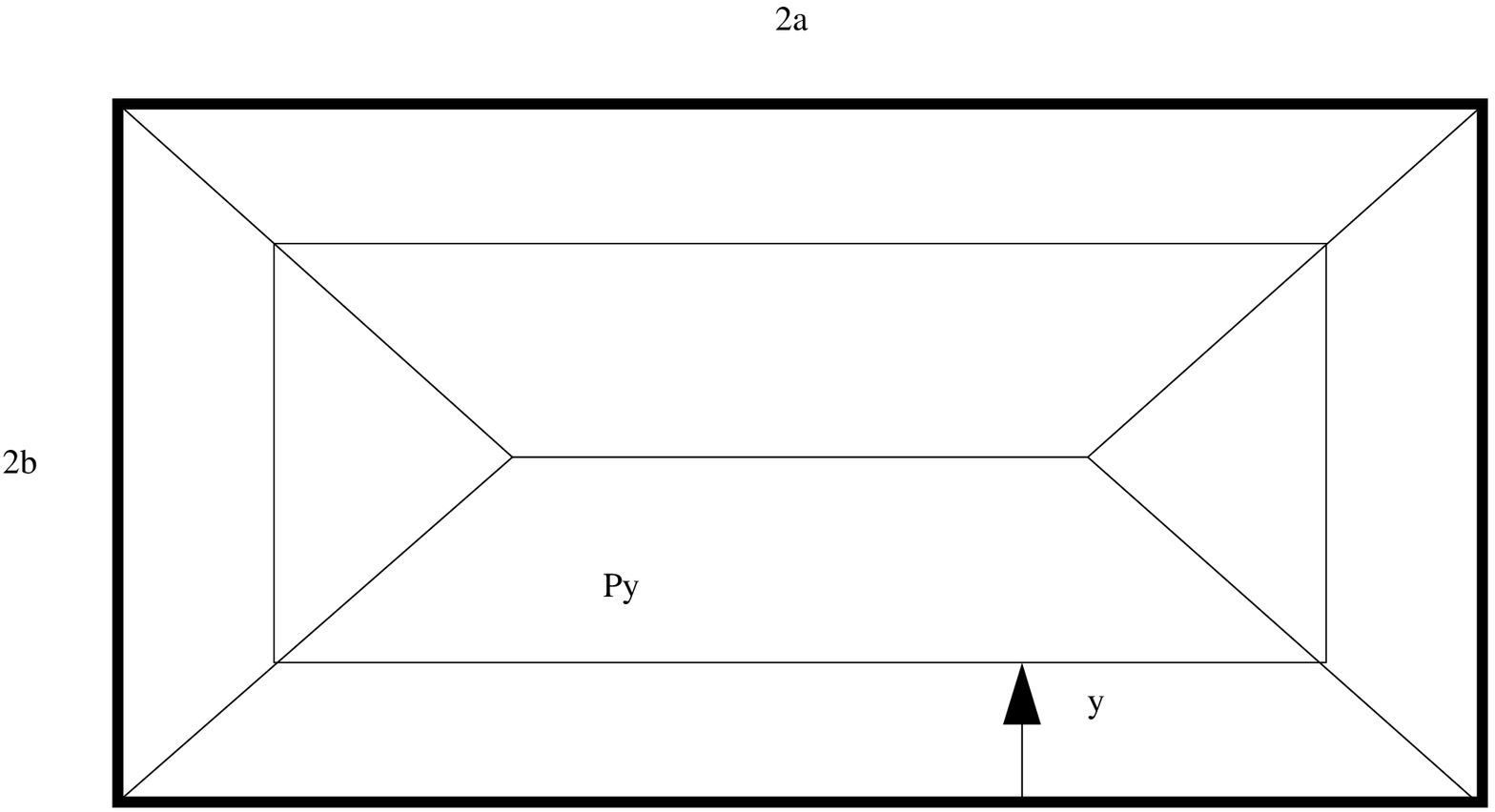} \hskip2.em
  \psfrag{de}[][][1.2]{$D_e/D_h$}
  \psfrag{e}[t][][1.2]{$1/\AR$}
  (b)
  \includegraphics[width=5.0cm,clip]{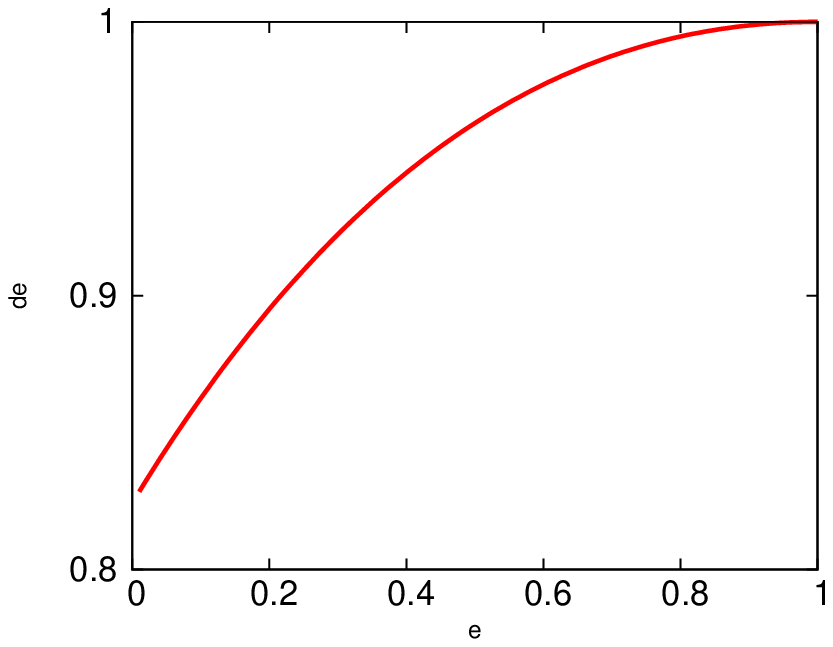}
\caption{Rectangular ducts: (a) cross section with indication of wall distance $y$, and associated area density, $P(y)$; (b) correction factor as a function of duct inverse aspect ratio.}
  \label{fig:rectangle}
 \end{center}
\end{figure}

Rectangular ducts have been extensively studied in the literature because of their 
wide range of use. Limited success of the classical hydraulic diameter concept
is recognized in this case, and corrections have been proposed, the most frequently 
used~\citep{jones_76} being based on the definition of a laminar-equivalent hydraulic diameter such that the friction coefficient in the laminar
regime is the same for ducts with any aspect ratio.
For a rectangular duct with sides length $a \ge b$, 
and aspect ratio $\AR = a/b$ (see figure~\ref{fig:rectangle}), the hydraulic diameter is $D_h=4 a b/(a+b)$,
the maximum wall distance is $y_m=b$,
the normalized perimeter function defined in 
equation~\eqref{eq:ub2} is given by
\begin{equation}
\tilde{P}= \frac {1+\AR-2 \eta}{\AR},
\end{equation}
and the geometric factor $\mathcal{C}$ defined in equation~\eqref{eq:ub3} is 
\begin{equation}
\mathcal{C} = -1-\frac 1{2 \AR}.
\end{equation}
It follows that the effective diameter is 
\begin{equation}
D_e = 2 b e^{\frac{\AR-1}{2 \AR}} = \frac {1+\AR}{2 \AR} D_h e^{\frac{\AR-1}{2 \AR}} , \label{eq:effdiam_rectangle}
\end{equation}
as graphically shown in figure~\ref{fig:rectangle}.
It should be noted that equation~\eqref{eq:effdiam_rectangle} predicts that the hydraulic diameter is the
correct length scale for square ducts, in line with DNS data at $\Rey_{\tau} \approx 1000$~\citep{pirozzoli_18}.

\begin{figure}
 \begin{center}
  \psfrag{f}[][][1.4]{$\lambda$}
  (a)
  \psfrag{Re}[][][1.4]{$\Rey_{D_h}$}
  \includegraphics[width=6.0cm,clip]{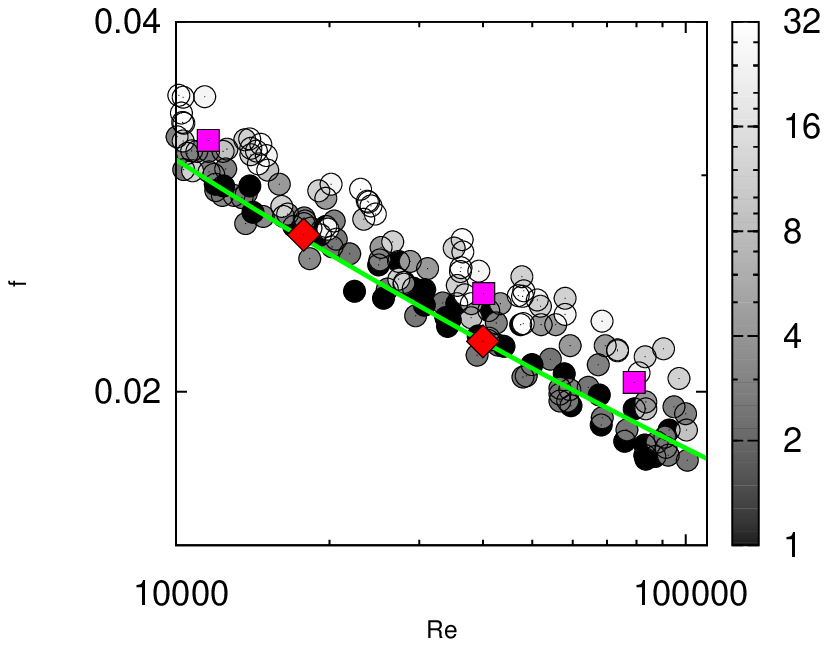}
  (b)
  \psfrag{Re}[][][1.4]{$\Rey_{D_e}$}
  \includegraphics[width=6.0cm,clip]{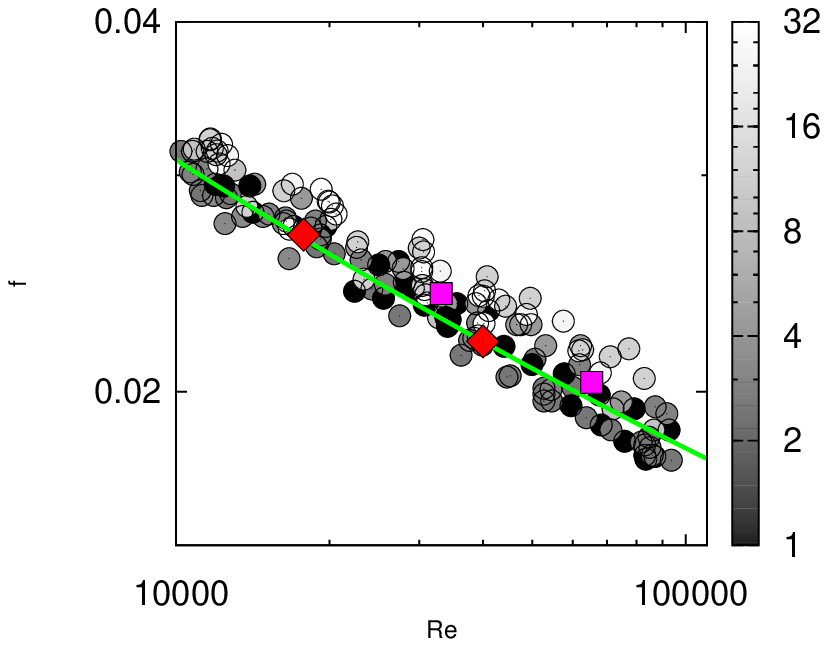}
\caption{Experimentally measured friction factor for rectangular ducts with various aspect ratios~\citep{jones_76}, as from color scale. 
Solid symbols denote DNS data for square duct~\citep[diamonds]{pirozzoli_18}, and plane channel flow~\citep[squares]{bernardini_14}.
Left, expressed as a function of the hydraulic diameter; right, expressed as a function of the 
effective diameter, equation~\ref{eq:effdiam_rectangle}. The solid line denotes the
K\'arm\'an-Prandtl friction law for a smooth circular pipe. The relative standard error is $8.1\%$ in panel (a), and $5.7\%$ in panel (b).}
  \label{fig:cf_rectangle}
 \end{center}
\end{figure}

Figure~\ref{fig:cf_rectangle} shows the measured friction coefficient in rectangular ducts with aspect 
ratios up to 31, as collected in the work of \citet{jones_76}, as a function of the
traditional hydraulic diameter (left panel) and as a function of the effective hydraulic diameter defined in
equation~\eqref{eq:effdiam_rectangle} (right panel). For reference, the K\'arm\'an-Prandtl friction law 
for smooth circular pipes
\begin{equation}
1/{\lambda}^{1/2} = 2 \log_{10} (\Rey_{D} {\lambda}^{1/2}) - 0.8, \label{eq:prandtl}
\end{equation}
is also shown, as well as DNS data for square ducts and plane channels.
As expected based on the previous discussion, use of the hydraulic diameter yields 
systematic underprediction of the friction factor for ducts with high aspect ratio.
This is particularly true for a subset of experimental data 
which even overshoot friction of plane channel flow,
thus raising doubts about their reliability.
Despite significant scatter among experimental data, it appears
that the log-law based effective diameter yields better universality of the distributions, especially
insofar as it tends to shift data points corresponding to high-aspect-ratio ducts closer
to the circular pipe case, including the infinite aspect ratio case. 
The remaining differences between plane channel and pipe flow, of about $2\%$, are likely due to differences in the
core velocity profiles, here neglected. In fact, the wake region in plane channel flow is weaker than in pipes,
hence the friction coefficient is higher, all the rest being the same.
Similar results as shown in figure~\ref{fig:cf_rectangle}(b) may be obtained from semi-empirical corrections~\citep[see][]{jones_76}, which however have very tenuous theoretical foundation.

\subsection{Annular ducts} \label{sec:annular}

\begin{figure}
 \begin{center}
  \psfrag{d}[][][1.2]{$y$}
  \psfrag{p}[][][1.2]{$X$}
  \psfrag{pw}[][][1.2]{$X_w$}
  \psfrag{D}[][][1.2]{$D$}
  \psfrag{L}[][][1.2]{$L$}
  \psfrag{Py}[][][1.2]{$P(y)$}
  \psfrag{R1}[][][1.2]{$R_1$}
  \psfrag{R2}[][][1.2]{$R_2$}
  (a)
  \includegraphics[width=4.0cm,clip]{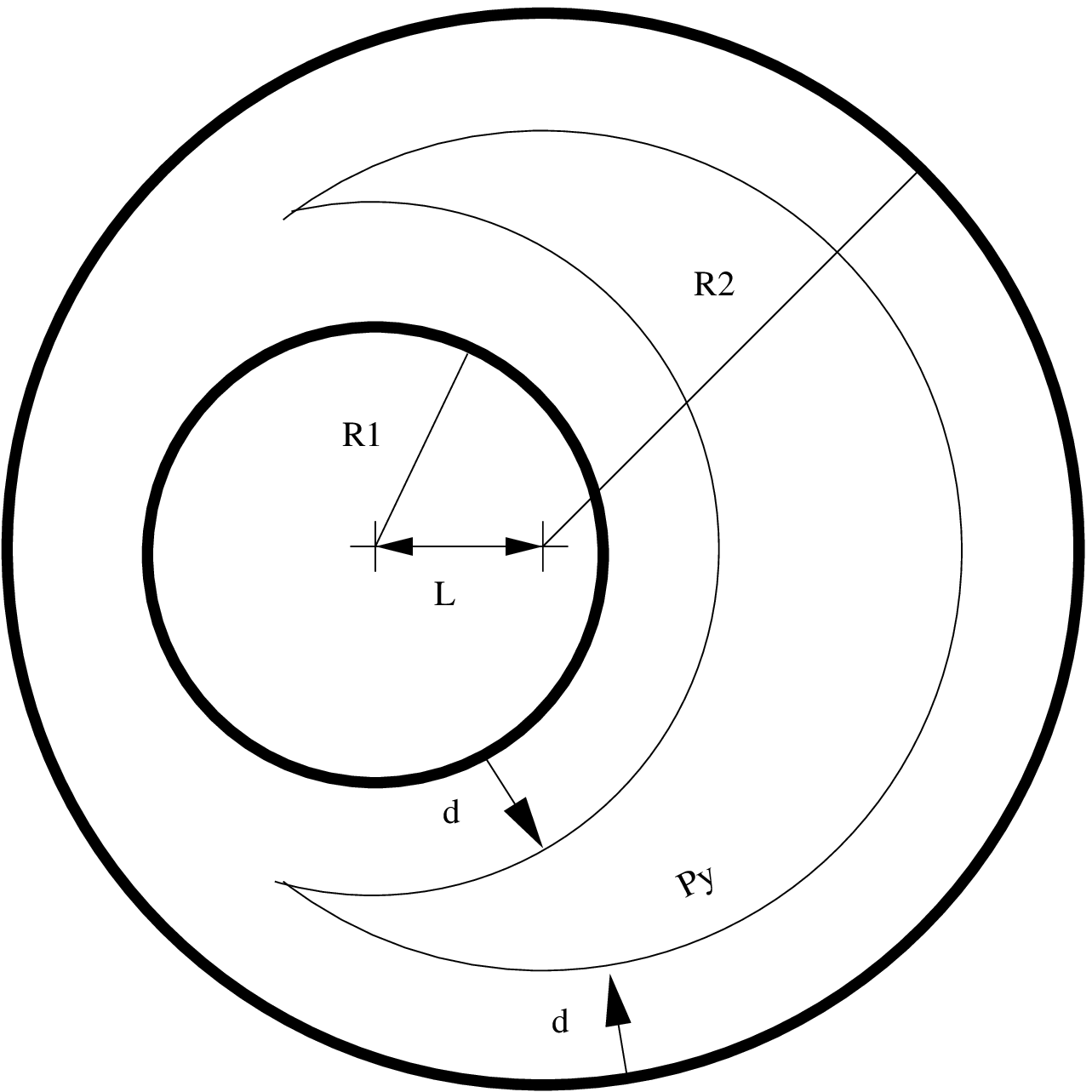} \hskip2.em
  \psfrag{de}[][][1.2]{$D_e/D_h$}
  \psfrag{e}[t][][1.2]{$\epsilon$}
  (b)
  \includegraphics[width=5.5cm,clip]{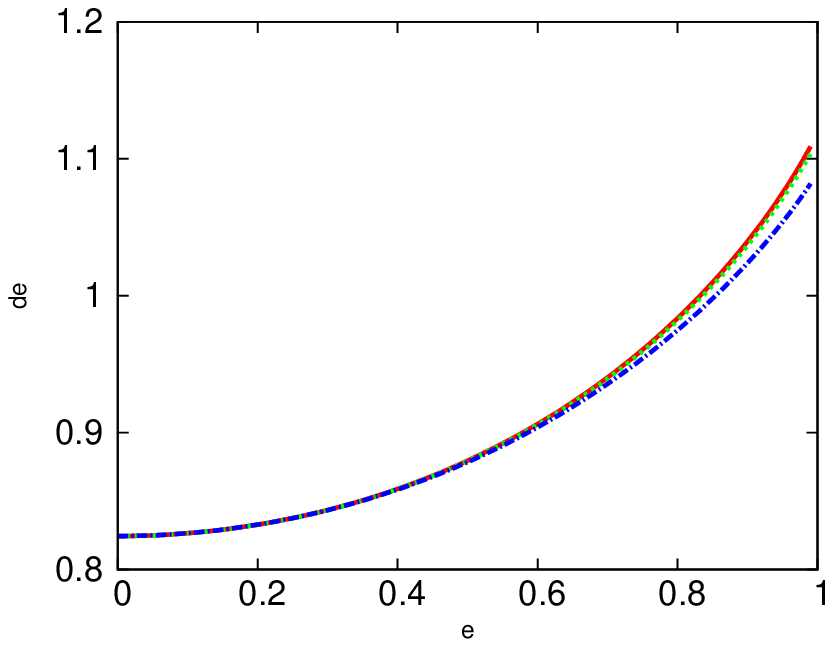}
  \caption{Annular ducts: (a) cross section with indication of wall distance $y$, and associated area density, $P(y)$, with $R_1$ and $R_2$ the radii of the inner and outer cylinder, respectively, and $L$ the offset between their centers; (b) correction factor as a function of the duct eccentricity $\epsilon$ for various duct diameters ratio: $\delta=1.33$ (solid), $\delta=1.78$ (dashed), $\delta=3.56$ (dot-dashed).}
  \label{fig:annular}
 \end{center}
\end{figure}

Flows in annular passages are common in mechanical engineering, and for instance they
are important in drilling wells, where mud passes through the drill shaft and the well casing
to remove cuttings and friction-generated heat. 
Let $L$ be the offset between the inner cylinder 
with radius $R_1$, and the outer cylinder with radius $R_2$, the geometry (see figure~\ref{fig:annular}(a)) 
is controlled by two parameters, namely the diameter ratio $\delta = R_2/R_1$, and the eccentricity, $\epsilon = L / (R_2-R_1)$.
The hydraulic diameter is simply $D_h = 2 (R_2-R_1)$, bearing no dependence on $\delta$ and $\epsilon$.
It is a simple matter to show that the maximum wall distance is in this case $y_m = 1/2 (R_2-R_1) (1+\epsilon)$, and the mean perimeter is
$\overline{P} = A/y_m = 2 \pi (R_2+R_1)/(1+\epsilon)$. After some algebra, the following expression is obtained for the 
normalized perimeter function
\begin{equation}
\tilde{P}(\eta) = \left\{ 
\begin{array}{lr}
        1+\epsilon ,   & \text{for } 0\leq \eta \leq \eta^* \\
        \frac{1+\epsilon}{\pi (\delta+1)} \left[ \delta \theta_2 + \theta_1 + \eta/2 (\delta-1)(1+\epsilon)(\theta_1-\theta_2) \right] , & \text{for } \eta^*\leq \eta \leq 1\\
        \end{array}
\right. , \label{eq:normper_annular}
\end{equation}
where $\eta^*=(1-\epsilon)/(1+\epsilon)$, 
\[ 
\cos \theta_1 = \frac{(\delta+1)\left[ (1+\epsilon) \eta - 1 \right] + \epsilon^2 (\delta-1)}{\epsilon (\delta-1) \left[ 2 + \eta (\delta-1)(1+\epsilon) \right]}, \quad
\cos \theta_2 = \frac{(\delta+1)\left[ (1+\epsilon) \eta - 1 \right] - \epsilon^2 (\delta-1)}{\epsilon (\delta-1) \left[ 2 \delta - \eta (\delta-1)(1+\epsilon) \right]}.
\quad 
\]

Numerical integration of function $\mathcal{C}$ defined in equation~\eqref{eq:ub3}, with $\tilde{P}$ given in equation~\eqref{eq:normper_annular} yields the result shown in figure~\ref{fig:annular}(b). As in periodic channel flow, the corrective factor over the hydraulic diameter is $\sqrt{e}/2$ at zero eccentricity, becoming closer and exceeding unity at increasing values of $\epsilon$. 
The dependence on the diameter ratio is quite weak, and confined to high values of $\epsilon$, a good fit for the data at $\epsilon \lesssim 0.6$ being $D_e/D_h = \sqrt{e}/2 + 0.217 \epsilon^2$.

\begin{figure}
 \begin{center}
  \psfrag{f}[][][1.2]{$\lambda$}
  (a)
  \psfrag{Re}[][][1.2]{$\Rey_{D_h}$}
  \includegraphics[width=6.0cm,clip]{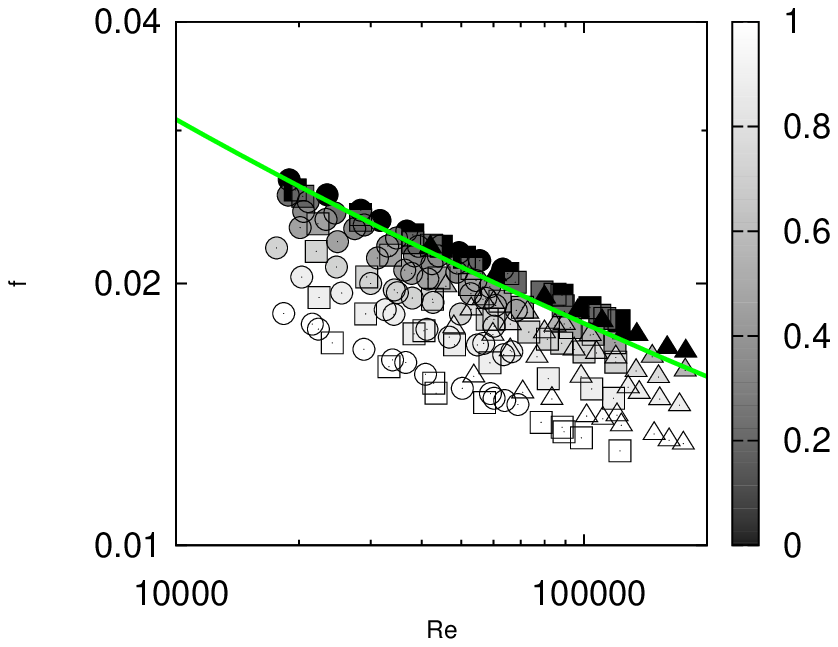}
  (b)
  \psfrag{Re}[][][1.2]{$\Rey_{D_e}$}
  \includegraphics[width=6.0cm,clip]{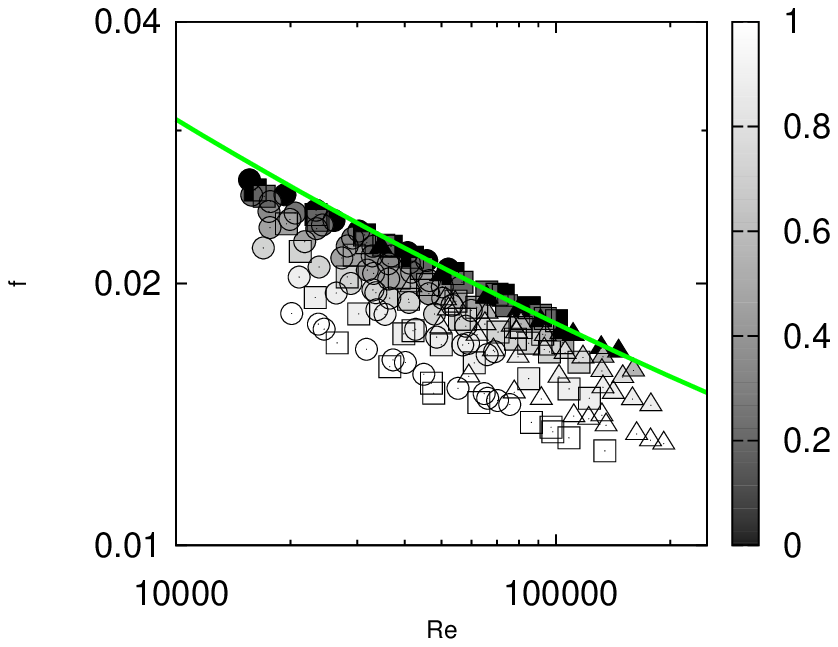}
\caption{Experimentally measured friction factor for annular ducts with different eccentricities~\citep{jonsson_66}, as from color scale, 
and different diameters ratio: $\delta=1.33$ (circles); $\delta=1.78$ (squares), $\delta=3.56$ (triangles).
Left, expressed as a function of the hydraulic diameter; right, expressed as a function of the 
effective diameter, equation~\ref{eq:effdiam}. The solid line denotes the
K\'arm\'an-Prandtl friction law for a smooth circular pipe. 
The relative standard error for $\epsilon \le 0.25$ is $3.0\%$ in panel (a), and $2.5\%$ in panel (b).}
  \label{fig:cf_annular}
 \end{center}
\end{figure}

Extensive measurements of concentric and eccentric pipe flows were carried out by \citet{jonsson_66},
in a wide range of duct eccentricities and diameters ratio.
Those authors found that the wall shear stress varies 
in the circumferential direction proportionally
to the duct eccentricity, with larger shear at the location of the widest gap.
Further difficulties are associated with the occurrence of locally laminar flow
at the smallest gap~\citep{nikitin_06}.
Comparison of the classical hydraulic diameter representation with predictions 
of equation~\eqref{eq:effdiam} are shown in figure~\ref{fig:cf_annular}.
The classical hydraulic diameter scaling (left panel) shows higher friction than given by 
the K\'arm\'an-Prandtl friction law at low duct eccentricity, and substantially lower 
at high eccentricity, 
mostly associated with the formation of regions of laminar flow.
The effective diameter is partially successful in shifting
the friction data for ducts with low eccentricity ($\epsilon \lesssim 0.25$, see color scale and figure caption) 
towards the universal distribution.
Although pointing in the right direction (recalling figure~\ref{fig:annular}(b)) the effective diameter
doesn't yield the same satisfactory behavior also at higher eccentricity, 
which is probably not unexpected as the underlying assumption of uniform wall stress is 
severely invalidated.

\subsection{Circular rod bundles} \label{sec:bundles}

\begin{figure}
 \begin{center}
  \psfrag{y}[][][1.0]{$y$}
  \psfrag{p}[][][1.2]{$X$}
  \psfrag{pw}[][][1.2]{$X_w$}
  \psfrag{D}[][][1.2]{$D$}
  \psfrag{L}[][][1.2]{$L$}
  \psfrag{Py}[][][1.0]{$P(y)$}
  \psfrag{R}[][][1.2]{$R$}
  (a)
  \includegraphics[width=4.5cm,clip]{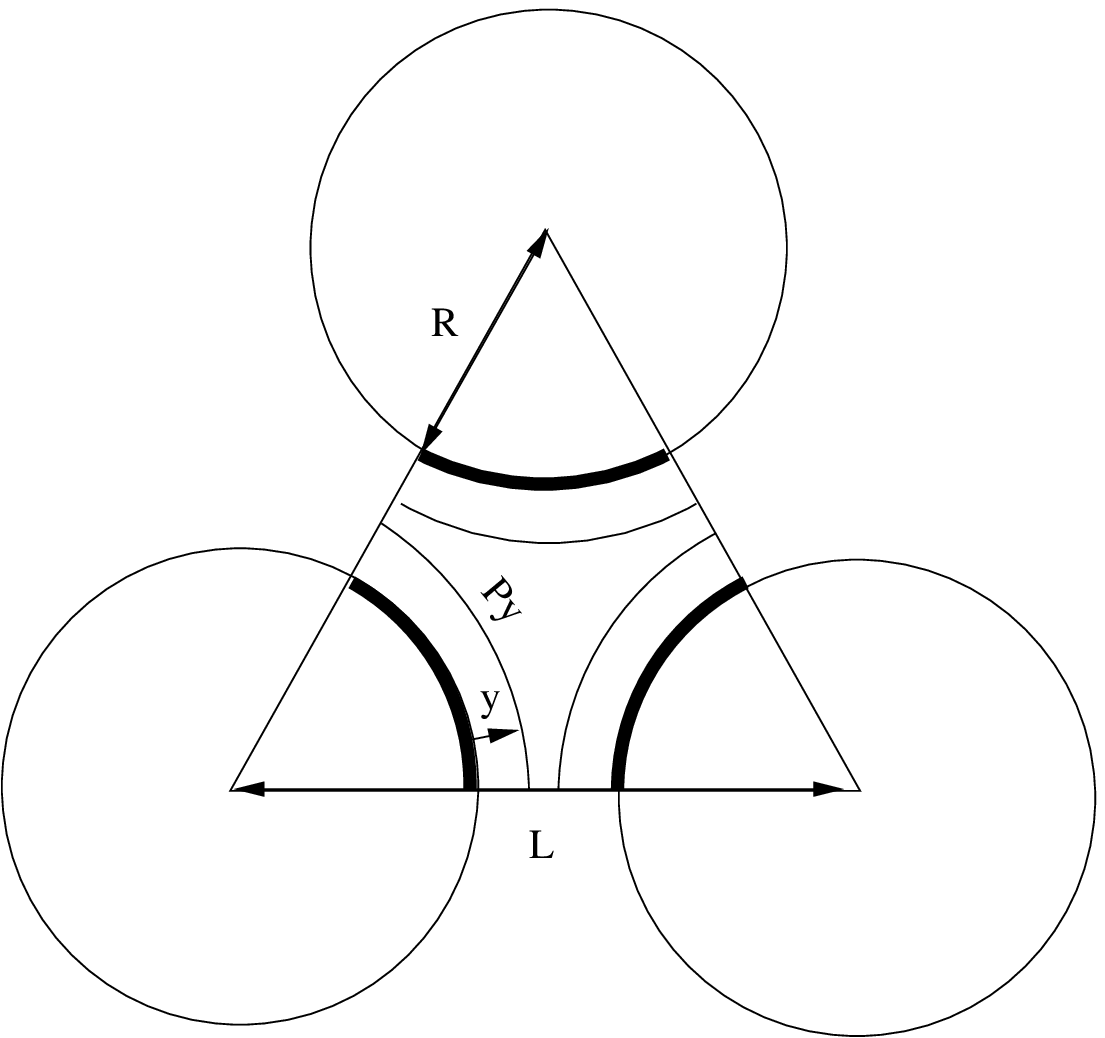} \hskip2.em
  \psfrag{de}[][][1.2]{$D_e/D_h$}
  \psfrag{e}[][][1.2]{$\epsilon$}
  (b)
  \includegraphics[width=5.2cm,clip]{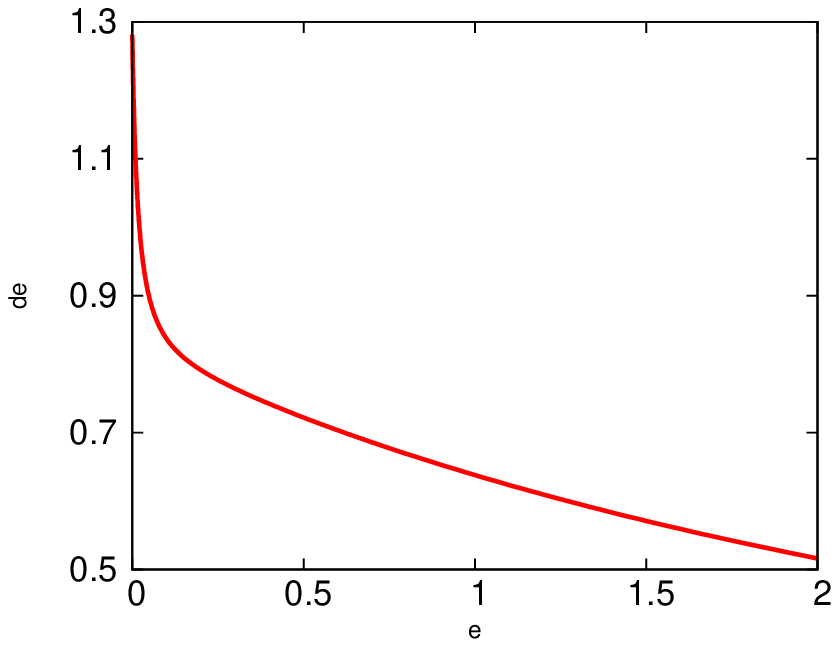}
  \caption{Rod bundles in hexagonal arrangement: (a) elementary module with indication of wall distance $y$, and associated area density, $P(y)$, with $R=D/2$ the rod radius, and $L$ the pitch between neighboring rods centers; (b) correction factor as a function of pitch parameter $\epsilon=L/D-1$.}
  \label{fig:bundle}
 \end{center}
\end{figure}

The prediction of friction in bundles of rods with circular cross section is important 
in the cooling channels of nuclear reactors, hence this flow has been extensively characterized in the past.
Although several arrangements of rods are possible, here we consider an hexagonal arrangement, under the assumption of nominally infinite number of rods, so as to neglect boundary effects related to the presence of  
confining walls. The relevant geometrical molecule for the study of this flow is shown in 
figure~\ref{fig:bundle}(a). The geometry of the typical cross section is identified through a 
single parameter, namely the pitch/diameter ratio, $L/D = 1+\epsilon$, $\epsilon=0$ corresponding to the limit case of tangent rods. The hydraulic diameter for the typical cross section is~\citep{rehme_73}
\[
D_h = D \, (1+\epsilon)^2 \frac {\tan{\alpha}}{\alpha},
\]
with $\alpha=\pi/6$. It turns out that the maximum wall distance is $y_m=R((1+\epsilon) \cos{\alpha} -1)$, the average
perimeter is 
\[
\overline{P} = \frac {3 \left[ (1+\epsilon)^2 \sin \alpha - \alpha \cos \alpha \right]}{(1+\epsilon)- \cos \alpha},
\]
and the normalized perimeter function is 
\begin{equation}
\tilde{P}(\eta) = \frac {6 R}{\overline{P}} \left\{ 
\begin{array}{lr}
        \alpha \left( 1 + \eta \, y_m / R \right),   & \text{for } 0\leq \eta \leq \eta^* \\
        \left( 1 + \eta \, y_m/R \right) \left[ \alpha - \arccos \left( \frac {1+\epsilon}{1+\eta \, y_m/R} \right) \right], & \text{for } \eta^*\leq \eta \leq 1\\
        \end{array}
\right. , \label{eq:normper_bundle}
\end{equation}
where $\eta^* = \epsilon R / y_m$.

Numerical integration of equation~\eqref{eq:ub3} yields the effective diameter as a function
of the parameter $\epsilon$, as shown in figure~\ref{fig:bundle}(b). The effective diameter is found to be 
greater than the hydraulic diameter for small $\epsilon$, with $D_e/D_h \to 1.281$ as $\epsilon \to 0$,
and decreasing as $1/\epsilon$ for large $\epsilon$.

\begin{figure}
 \begin{center}
  \psfrag{f}[][][1.1]{$\lambda$}
  (a)
  \psfrag{Re}[][][1.1]{$\Rey_{D_h}$}
  \includegraphics[width=6.0cm,clip]{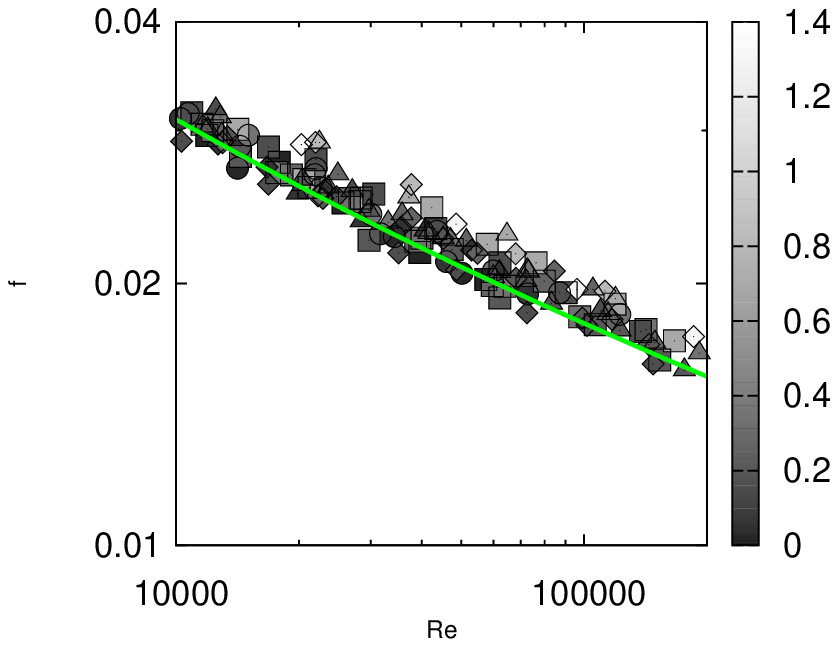}
  (b)
  \psfrag{Re}[][][1.1]{$\Rey_{D_e}$}
  \includegraphics[width=6.0cm,clip]{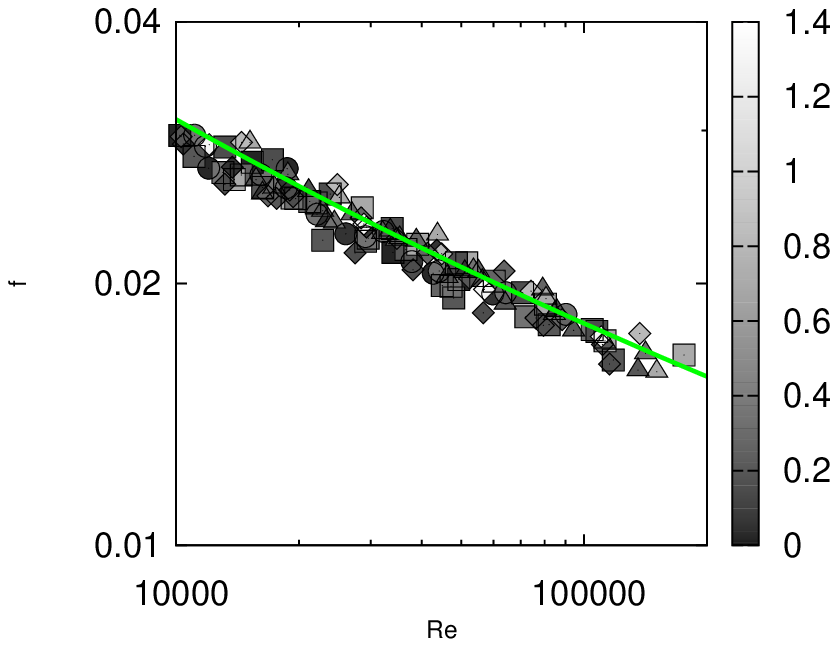}
\caption{Experimentally measured friction factor for rod bundles for different pitch parameters $\epsilon$~\citep{rehme_73}, as from color scale, 
and different numbers of rods: $N=61$ (circles); $N=37$ (squares), $N=19$ (triangles), $N=7$ (diamonds).
Left, expressed as a function of the hydraulic diameter; right, expressed as a function of the 
effective diameter, equation~\ref{eq:effdiam}. The solid line denotes the
K\'arm\'an-Prandtl friction law for a smooth circular pipe.
The relative standard error is $6.8\%$ in panel (a), and $4.5\%$ in panel (b).}
  \label{fig:cf_bundle}
 \end{center}
\end{figure}

In figure~\ref{fig:cf_bundle} we show experimental data by \citet{rehme_73} in the 
range of pitch parameters $0.025 \le \epsilon \le 1.320$. When normalized with respect to the hydraulic 
diameter (left panel), the data show consistently higher friction than expected, the higher is $\epsilon$.
The effective diameter (right panel) does eliminate most
large overshoots, reestablishing the universal friction law for all data at sufficiently high pitch ratio (light shades).
Data points corresponding to small values of $\epsilon$ are overestimated, presumably because flow 
becomes locally laminar in the presence of narrow gaps, an effect which again cannot be 
captured given the initial assumptions.

\section{Conclusions} \label{sec:concl}

We have derived predictive formulas for friction in ducts with complex shape, 
under the assumption that the friction distribution along the duct perimeter is uniform, and 
that the velocity at any given point in the cross section is controlled by the nearest wall
through an assumed inner-layer log law, upon neglect of core deviations.
The leading conclusion is that the classical hydraulic diameter is the proper length
for many common duct shapes such as triangles and regular polygons,
thus providing theoretical support to its widespread use.
This finding is supported by a number of recent DNS in polygonal ducts which with very good 
precision show collapse of friction data on the universal K\'arm\'an-Prandtl distribution.
A second important conclusion is that deviations from the classical
hydraulic diameter scaling should arise in more general duct shapes, for which
the effective diameter defined in equation~\eqref{eq:effdiam} is expected to be a more accurate choice.
The latter can be easily evaluated, either analytically or numerically,
based on the duct cross-sectional shape. Differences with respect to friction predictions 
based on the classical hydraulic diameter
are generally small, but more sensible as the duct aspect ratio is much different than unity,
amounting in practical terms to a few percent.
Re-evaluation of classical experiments in smooth ducts with moderately complex shape, 
namely rectangular and annular ducts, and circular rod bundles, supports 
small but consistent predictive improvements when the effective diameter
is used instead of the traditional hydraulic diameter, 
with all due caveats incurred with tracing differences of a 
few percent within scattered experimental data. 
Of course, given the assumptions made in the derivation, 
the log-law based effective diameter does not perform well in cases
in which the wall shear stress is far from uniform, such as ducts with narrow gaps 
or acute angles, which may even feature locally laminar flow. 
Measurements at higher Reynolds number and/or for rough ducts would be desirable, 
to be able to more clearly ascertain the predictive power of the log-law based effective diameter, 
and to compare with existing correlations based on extrapolation of laminar results.

\begin{acknowledgments}
This paper has benefited from exchange of ideas with P. Orlandi and P.R. Spalart.
\end{acknowledgments}
\bibliographystyle{jfm}
\bibliography{references}
\end{document}